\documentclass[fleqn,usenatbib]{mnras}

\usepackage{newtxtext,newtxmath}

\usepackage[T1]{fontenc}
\usepackage{ae,aecompl}

\usepackage{graphicx}	
\usepackage{amsmath}	
\usepackage{amssymb}	
\usepackage{caption}

\newcommand{\avenf}{\bar{x}_{\ion{H}{I}}}
\newcommand{\aveTs}{\bar{T}_{\rm S}}

\title[Spin-temperature dependence of the 21cm -- LAE cross-correlation]{The spin-temperature dependence of the 21cm -- LAE cross-correlation}

\author[Heneka \& Mesinger]{
Caroline Heneka\thanks{E-mail: caroline.heneka@sns.it} \&
Andrei Mesinger
\\
Scuola Normale Superiore, Piazza dei Cavalieri 7, 56126 Pisa, Italy
}

\date{Accepted XXX. Received YYY; in original form ZZZ}

\pubyear{2020}

\begin{document}
\label{firstpage}
\pagerange{\pageref{firstpage}--\pageref{lastpage}}
\maketitle

\begin{abstract}
Cross-correlating 21cm with known cosmic signals will be invaluable proof of the cosmic origin of the first 21cm detections.  
As some of the widest fields available, comprising thousands of sources with reasonably known redshifts, narrow-band Lyman alpha emitter (LAE) surveys are an obvious choice for such cross-correlation.
Here we revisit the 21cm -- LAE cross-correlation, relaxing the common assumption of reionization occurring in a pre-heated intergalactic medium (IGM). 
Using specifications from the Square Kilometre Array and the Subary Hyper Supreme-Cam, we  present new forecasts of the 21cm -- LAE cross-correlation function at $z\sim7$. 
We sample a broad parameter space of the mean IGM neutral fraction and spin temperature, ($\avenf$, $\aveTs$).
The sign of the cross-correlation roughly follows the sign of the 21cm signal: ionized regions which surround LAEs correspond to relative hot spots in the 21cm signal when the neutral IGM is colder than the CMB, and relative cold spots when the neutral IGM is hotter than the CMB.  The amplitude of the cross-correlation function generally increases with increasing $\avenf$, following the increasing bias of the cosmic HII regions.  As is the case for 21cm, the strongest cross signal occurs when the IGM is colder than the CMB, providing a large contrast between the neutral regions and the ionized regions which host LAEs.
We also vary the topology of reionization and the epoch of X-ray heating.  The cross-correlation during the first half of reionization is sensitive to these topologies, and could thus be used to constrain them.
\end{abstract}




\section{Introduction}

The hyperfine transition of $\ion{H}{I}$, releasing a photon with a rest-frame wavelength of 21-cm, can revolutionize our understanding of the early Universe.  Current radio interferometers such as the Low Frequency Array~\citep[LoFAR]{vanHaarlem13},\footnote{\url{http://www.lofar.org/}} and the Murchison Wide-field Array~\citep[MWA]{Tingay13}\footnote{\url{http://www.mwatelescope.org/}} are trying to statistically detect the cosmic 21-cm signal via redshift evolution of the 21-cm fluctuations.

The cosmic 21-cm signal is commonly expressed in terms of the offset of the 21-cm brightness temperature, $\delta T_{\rm b}(\nu)$, relative to the temperature of the cosmic microwave background (CMB), $T_{\rm CMB}$ \citep[e.g.][]{FOB06}:
\begin{equation}\label{eq:delT}
\begin{split}
\delta T_{\rm b} &\approx 27 x_{\ion{H}{I}} (1 + \delta_{\rm nl}) \left(\frac{H}{{\rm d}v_{\rm r}/{\rm d}r + H}\right)\left(1 - \frac{T_{\rm CMB}}{T_{\rm S}}\right)\\
                 &\quad \times \left( \frac{1+z}{10} \frac{0.15}{\Omega{\rm m}h^2} \right)^{1/2} \left(\frac{\Omega_{\rm b}h^2}{0.023} \right),
\end{split}
\end{equation}
where $ x_{\ion{H}{I}}$ is the neutral fraction, $T_{\rm S}$ is the gas spin temperature, $\delta_{\rm nl} \equiv \rho/{\bar \rho}-1$ is the gas overdensity, $H(z)$ is the Hubble parameter, ${\rm d}v_{\rm r}/{\rm d}r$ is the gradient of the line-of-sight component of the velocity and all quantities are evaluated at redshift $z=\nu_0/\nu - 1$, where $\nu_0$ is the 21-cm frequency.   As can be seen from eq. (\ref{eq:delT}), the 21-cm signal is sensitive to both the thermal and ionization state of the intergalactic medium (IGM), which are likely determined by the UV and X-ray emission of the first galaxies.  Therefore, the timing and structure of the 21-cm signal can indirectly inform us about the properties of galaxies which will remain undetected in the foreseeable future (e.g. \citealt{OShea15, DF18}), besides informing us about underlying cosmology and structure formation (e.g. \citealt{2013JCAP...01..003B,2018JCAP...10..004H,2019arXiv191002763L}).

However, unlocking this treasure trove will be a long and difficult journey.  We need to dig out the signal from underneath foregrounds and systematics that are many orders of magnitude stronger. As part of this effort, it is imperative to have a sanity check to test that the recovered 21-cm signal is genuinely cosmological. Cross-correlation with confirmed high-$z$ sources is ideal for this task (e.g. \citealt{Lidz09, Park14, Vrbanec16, SMG16, HTD18, BL18, Moriwaki19, PRA19}), as is cross-correlation with maps of emission line fluctuations (e.g. \citealt{2017ApJ...848...52H,2019BAAS...51g..23C,2019BAAS...51c.282C}).  Since cross-correlation is less sensitive to foreground contamination, it could also be used to estimate the general evolution of the 21cm auto-power spectrum, providing a valuable cross-check to the auto power estimates (e.g. \citealt{BL18, BVL19}).

Currently, narrow-band selected Lyman alpha emitting galaxies (LAEs) are the most promising candidates for such a cross-correlation.  Galaxies preferentially reside inside the large-scale overdensities which are the first to ionize; therefore, one would expect a galaxy field (or indeed any field which traces matter) to anti-correlate with 21-cm during the epoch of reionization (EoR) (assuming $\aveTs \gg T_{\rm CMB}$ in eq. 1; e.g. \citealt{Lidz07}).
Wide-field, narrow-band surveys, such as those from the Subaru telescope\footnote{\url{https://hsc.mtk.nao.ac.jp/ssp/}} \citep{Ouchi10, Konno14, Ouchi18, Konno18}, can provide thousands of $z\sim$7 LAEs on large scales with reasonably well-known redshifts (localized to $\Delta z \sim 0.1$).  Although most of the transverse (on sky) modes of the LAE maps correspond to those that in 21-cm are expected to be dominated by foregrounds, the LAE - 21cm cross-correlation might be detectable with first generation instruments, under optimistic assumptions (e.g. \citealt{Lidz09, Vrbanec16, SMG16}).  The SKA-low phase 1 will be able to detect the cross-correlation in just a few hours, even under more pessimistic assumptions about foreground contamination and the topology of reionization (e.g. \citealt{SMG16, 2017ApJ...836..176H, HTD18, Kubota19, Vrbanec20}).  This makes the LAE-21cm cross-correlation an ideal sanity check for data processing pipelines currently under development.

However, previous estimates of the LAE-21cm cross-correlation during the EoR (for surveys at $z\sim7$) made the simplifying assumption that the IGM was already pre-heated to $T_S \gg T_{\rm CMB}$ before reionization.  Under this simplifying assumption, the temperature term drops out from eq. (1).
However, subsequent 21-cm forecasts calibrated to high-$z$ luminosity functions (LFs) suggest that this is unlikely to be the case (e.g. \citealt{MFS17, Park19}).
High-$z$ galaxy observations from {\it Hubble} (e.g. \citealt{Finkelstein15, Atek15, Bouwens15, Oesch18, Ishigaki18, Bhatawdekar19}) suggests that the star formation rate density (SFRD) decreases beyond $z>$ 6--10 quicker than previously assumed.  Since the dominant source of IGM heating at these redshifts are expected to be X-rays from high-mass X-ray binaries (HMXBs; e.g. \citealt{Fragos12, Pacucci14, Lehmer16}), whose luminosities scale with the SFRD (e.g. \citealt{Lehmer10, MGS12_HMXB, Fragos13, BKP14, Douna15, Lehmer16}), the dropping SFRD implies that it is unlikely the IGM has been significantly pre-heated before the EoR (e.g \citealt{Das17, MFS17, MF17, Eide18, Park19}).
\footnote{This would not be true if the recent claim of a detection of a global 21-cm absorption signal at $z\sim17$ by EDGES \citep{Bowman18} is confirmed.  If this signal is indeed cosmological, it would require the IGM to have already been heated well before the EoR (e.g. \citealt{EW18, FB19}), likely by HMXBs residing in a unique population of faint, unseen galaxies (e.g. \citealt{MF19, MMF19, Qin20}).  In this scenario, it would be safe to assume $T_S \gg T_{\rm CMB}$ when computing the LAE-21cm cross-correlation at $z\sim$7.  However, the interpretation of the EDGES detection as having a cosmic origin currently remains quite controversial (e.g. \citealt{Hills18, Bradley18, SP19}).}

In this paper we relax the standard assumption of a pre-heated IGM, allowing the IGM temperature to vary when calculating the LAE-21cm cross-correlation.
We use 3D semi-numerical simulations which self-consistently compute the thermal and ionization evolution of a multi-phase IGM, taking two extremes for which halos host the dominant sources of X-ray and UV photons.  From these, we create mock survey realizations for SKA1-low and Subaru HSC, presenting the corresponding LAE-21cm cross-correlation as a function of the mean IGM neutral fraction, $\bar{x}_{\ion{H}{I}}$, and the mean spin temperature of the neutral IGM, $\aveTs$.\footnote{Throughout this work, $\aveTs$ refers to the HI volume weighted mean spin temperature; i.e. corresponding to the average spin temperature in the {\it neutral} IGM.}

The paper is organized as follows.  In \S \ref{sec:sim} we introduce the reionization simulations and modelling of LAEs employed in this study, as well as mock survey realizations. In \S~\ref{sec:cross} we present the resulting cross-correlation signal and its model dependencies.  Finally, we present our conclusions in \S~\ref{sec:end}.

\section{Methods}\label{sec:sim}

Here we describe the large-scale simulations needed for computing both the 21cm brightness temperature and the IGM attenuation of Ly$\alpha$ emerging from galaxies.  We then discuss our empirical LAE model, calibrated to reproduce luminosity functions and post-reionization clustering.  Finally, we describe how we generate mock SKA1-LOW and Subaru LAE surveys, and how we compute their cross-correlation.

\subsection{Reionization and Cosmic Dawn simulations}
\label{sec:21cm} 

In this study we use results from the Evolution of 21cm Structure (EOS) project released in~\citet{Mesinger16_EOS},\footnote{\url{http://homepage.sns.it/mesinger/EOS.html}}.  These have a 1.6 Gpc box length, and are computed on a 1024$^3$ grid, comprising the largest public 21cm simulation of the EoR. The EOS simulations were created with 21cmFASTv2~\citep{SM14}, which includes sub-grid prescriptions for inhomogeneous recombinations as well as photo-heating suppression of the gas fraction in small halos.  The X-ray emissivity of galaxies, which determines the inhomogeneous evolution of IGM temperature pre-reionization, is calibrated to match HMXB observations of local star forming galaxies \citep{MGS12_ISM}. The simulations also self-consistently compute the Lyman series radiation background,
that determines how closely the spin temperature tracks the gas kinetic temperature through Wouthuysen-Field (WF) coupling \citep{Wouthuysen52, Field58}.  However, over most of the parameter range of interest here, the spin temperature is already closely coupled to the gas kinetic temperature.

Assuming $T_S \gg T_{\rm CMB}$, the dominant uncertainty in determining the observable 21cm - LAE cross-correlation at a given $\bar{x}_{\ion{H}{I}}$ is the EoR morphology, with the LAE prescription only affecting small scales \citep{SMG16, Kubota19}.
We therefore take the two extreme models for the EoR morphology presented in EOS: (i) a faint galaxy model characterized by many small HII regions (SmallHII), and a (ii) bright galaxy model characterized by fewer, larger HII regions (LargeHII).  These are differentiated by different star-formation prescriptions, corresponding to efficient star formation in either faint or bright galaxies.  In both cases, the ionizing escape fraction is calibrated to yield similar Thompson scattering optical depths, consistent with estimates from {\it Planck} \citep{Planck16}.

In this work we want to quantify how the 21cm -- LAE cross-correlation depends on the mean IGM neutral fraction and spin temperature.  Unfortunately, it would be too costly to run multiple ultra-large scale EOS simulations, varying the source prescriptions.  Instead, we follow the common approach of adjusting the redshifts of the component maps in order to obtain the 21cm brightness field at $z=6.6$, corresponding to the Subary narrow-band survey (e.g. \citealt{McQuinn07, Jensen14, SMG16, Mason17}).  Specifically, we take the EOS ionization and spin temperature coeval boxes corresponding to given values of $\bar{x}_{\ion{H}{I}}$ and $\bar{T}_S$, combining them with the $z=6.6$ density field, in order to compute the brightness temperature that is to be cross-correlated with mock LAE maps.  This adjustment of the (unknown) timing of the EoR and Epoch of Heating (EoH), is roughly analogous to  adjusting parameters for the emission of ionizing and X-ray photons (such as the ionizing escape fraction and the X-ray luminosity; e.g. \citealt{McQuinn07, GM17_21CMMC}).

The EOS fiducial cosmology corresponds to the best-fit parameter values from \citet{Planck15}, with $\left(\Omega_{\Lambda},\Omega_\mathrm{m},\Omega_\mathrm{b},n_\mathrm{s},\sigma_8,H_0\right)=\left( 0.69, 0.31, 0.048,0.97,0.82,68 \,\mathrm{km}\,\mathrm{s}^{-1}\mathrm{Mpc}^{-1}\right)$ for a flat $\Lambda$CDM cosmology. We use these cosmological parameters throughout this paper. Unless stated otherwise, all quantities are given in comoving units.

\subsection{The LAE signal} \label{sec:LAE}

To assign LAEs to host halos we follow the procedure presented in~\citet{SMG16}, which we briefly summarize here.

A galaxy's observed Ly$\alpha$ luminosity, $L_\alpha$ is connected to the intrinsic luminosity which escapes into the IGM following radiative transfer through the dusty, multi-phase ISM (e.g. \citealt{Gronke17, Behrens19}), $L_\alpha^{\mathrm{intr}}$, via:
\begin{equation}
L_\alpha = L_\alpha^{\mathrm{intr}} e^{-\tau_\mathrm{Ly\alpha}} ,
\end{equation}
where $\tau_\mathrm{Ly\alpha}$ denotes the IGM optical depth along the line of sight (LOS). The IGM optical depth is computed by tracing through the HI density and velocity fields of the EOS 21cm simulations along a chosen LOS direction.  In principle, one should assume an emerging Lyman alpha emission line profile, and integrate the frequency dependent IGM optical depth over this profile in order to determine the observed Lyman alpha luminosity.  For simplicity, and since we do not really know the intrinsic line profile, we evaluate the IGM absorption at a fixed velocity offset $\Delta v \approx$ -230 km s$^{-1}$ redward of the systemic redshift, consistent with current observations of the typical velocity shift of the Ly$\alpha$ line from galaxies (e.g. \citealt{Shibuya14, Stark15, Sobral15, Hoag19}).  As shown in Appendix A of \citet{SM15}, the clustering properties of the observed narrow-band selected LAEs are insensitive to the intrinsic profile, {\it provided that they are normalized to a fixed observed number density}, since it is degenerate with the intrinsic narrow-band luminosity discussed below (see also \citealt{Jensen14}).

The intrinsic luminosity is related to the host halo mass $M_\mathrm{h}$ with
\begin{equation}
L_\alpha^{\mathrm{intr}} = L_\alpha^{\mathrm{min}} \left( \frac{M_\mathrm{h}}{M_\alpha^{\mathrm{min}} } \right)^\beta \chi , 
\end{equation}
where  $L_\alpha^{\mathrm{min}}$ is the minimum observed Ly$\alpha$ luminosity at a corresponding halo mass $M_\alpha^{\mathrm{min}}$, $\beta$ is the slope of the relation and the random variable $\chi$ accounts for the stochasticity of Lyman alpha emission, having a probability $f_\mathrm{duty}$ to be unity and (1-$f_\mathrm{duty}$) to be zero.  We take $\beta =1$, consistent with findings for $z\sim 4$ LAEs~\citep{Gronke15}.

For every choice of $\bar{x}_{\rm HI}(z=6.6)$, we vary $f_\mathrm{duty}$ together with $M_\alpha^{\mathrm{min}}$ to match the observed $z=6.6$ LAE number density of $\bar{n}_\mathrm{LAE}\sim 5\times 10^{-4}\mathrm{Mpc}^{-3}$ for the limiting magnitude of $L_\alpha^{\mathrm{min}} = 2.5\times 10^{42}$erg$\,$s$^{-1}$, as found in the Subaru SupremeCam ultra-deep (UD) field \citep{Ouchi10}.  The HSC ultra-deep field will have the same limiting magnitude, as we discuss below.  For reference, in the case of $\bar{x}_{\rm HI}(z=6.6) \approx 0$, this procedure results in an average halo mass of $\bar{M}_\mathrm{h}\approx 2\times 10^{10}M_{\odot}$ for $M_\alpha^{\mathrm{min}}\sim 8\times 10^{9}M_{\odot}$ and a duty cycle of $f_\mathrm{duty}\approx 0.02$. We  use the calibration discussed here throughout the paper together with the LargeHII and SmallHII reionization models.  As shown in Figures 1 and 2 in \citet{SM15},  this procedure results in narrow-band selected LAEs whose luminosity functions and clustering properties are consistent with observations.

\subsection{Survey realisations}\label{sec:mock}
\subsubsection{Mock LAE catalogue: HSC ultra-deep field}\label{sec:mockLAE}

Our fiducial LAE observation is a mock Subaru/HSC UD-like survey at $z\sim 6.6$ with a survey area of $\sim$ 3.5 deg$^2$, a systemic redshift uncertainty of $\Delta z =0.1$ (corresponding to a slice of roughly 38$\,$Mpc at a redshift of $z=6.6$), and a limiting narrow-band luminosity of $L_\alpha^{\mathrm{min}} = 2.5\times 10^{42}$erg$\,$s$^{-1}$ (M. Ouchi, private communication). The LAE maps computed as described in the previous section, are thus cut into slices of width 38$\,$Mpc and we take a patch of 3.5 deg$^2$ in order to obtain the 2D distribution of LAEs for the chosen field. We select different non-overlapping fields from our simulation box when calculating the scatter from cosmic variance (see \S~\ref{sec:rcorr}).

\begin{figure*}
\centering 
\includegraphics[width=\columnwidth]{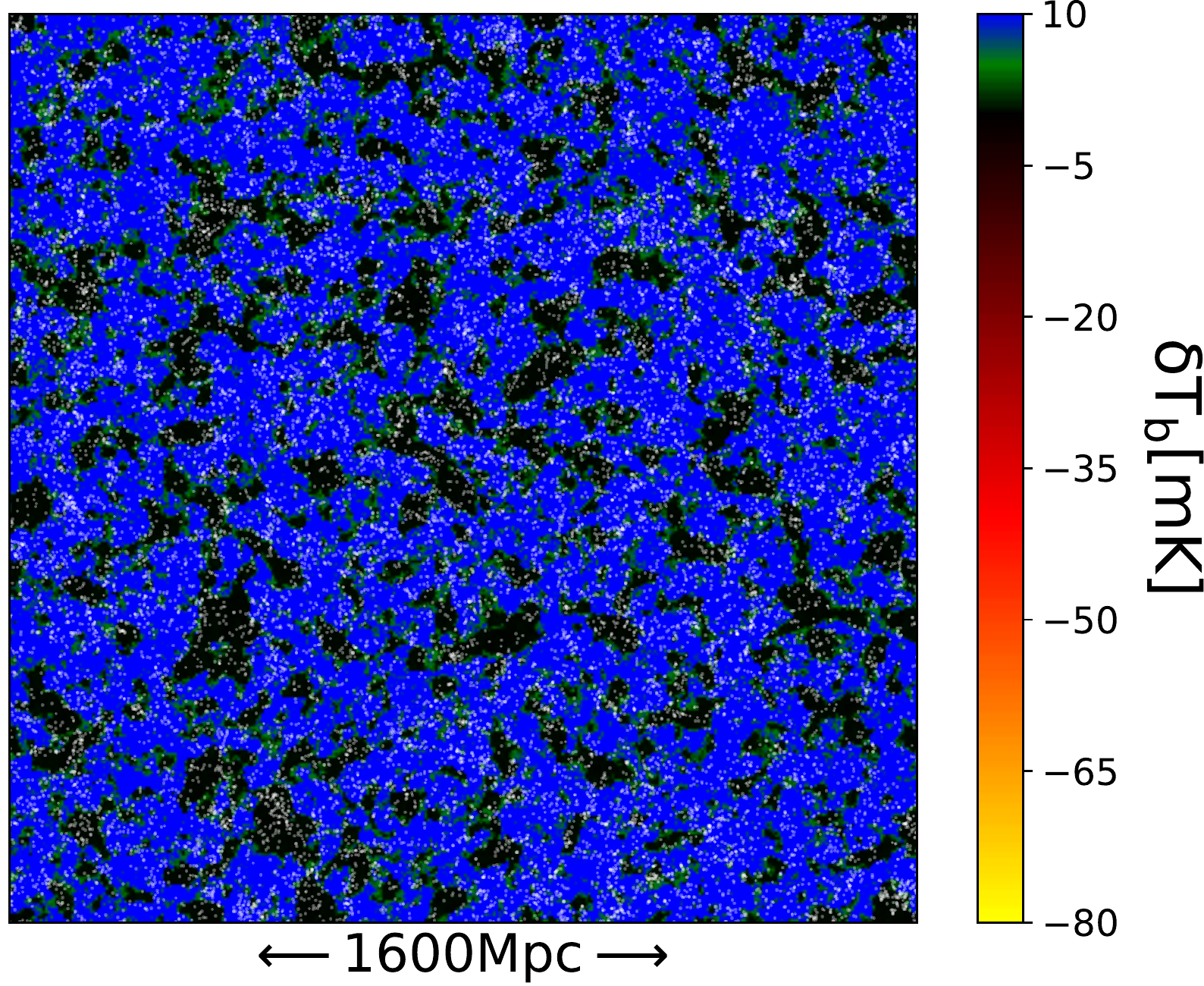}
\includegraphics[width=\columnwidth]{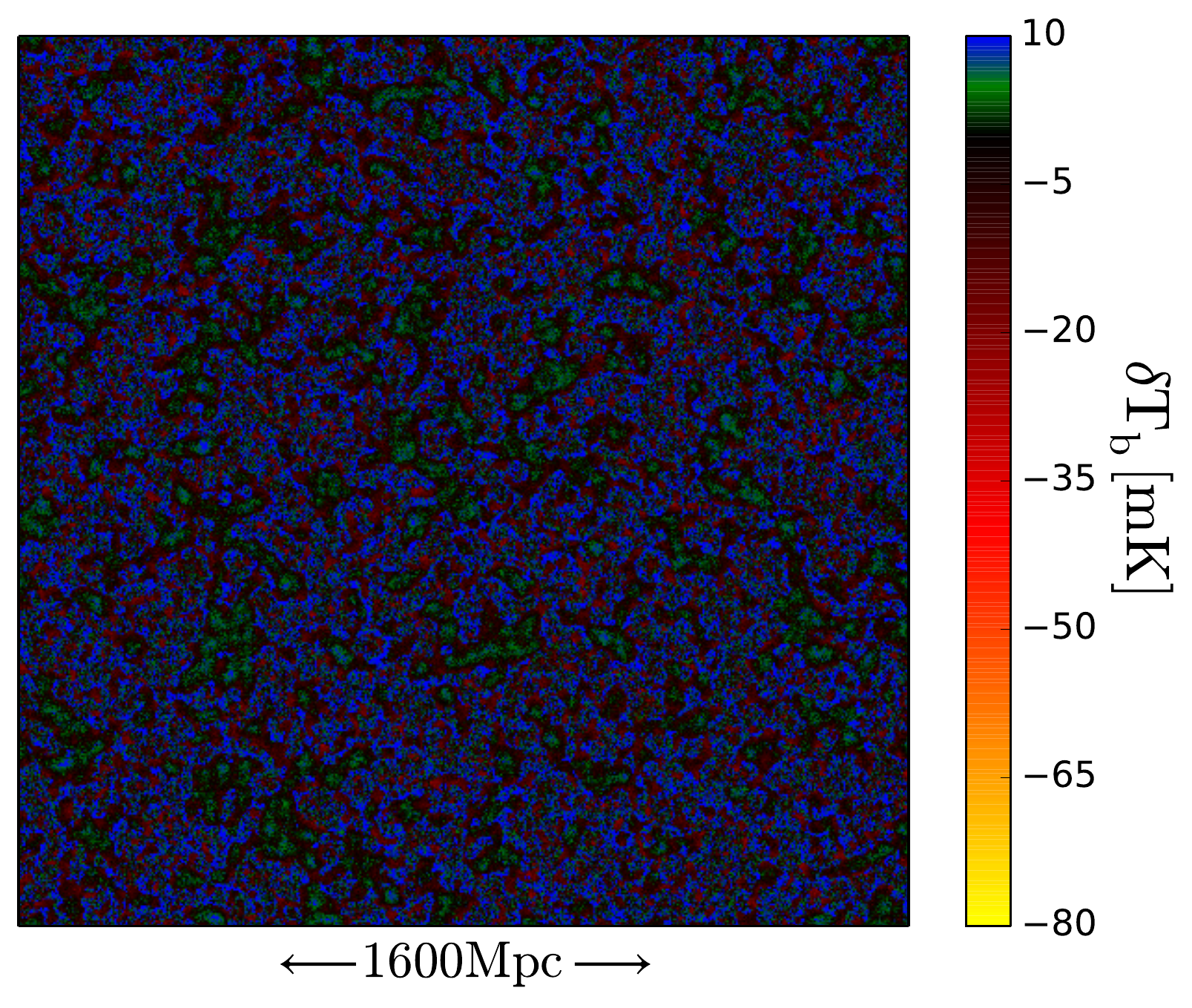}
\includegraphics[width=\columnwidth]{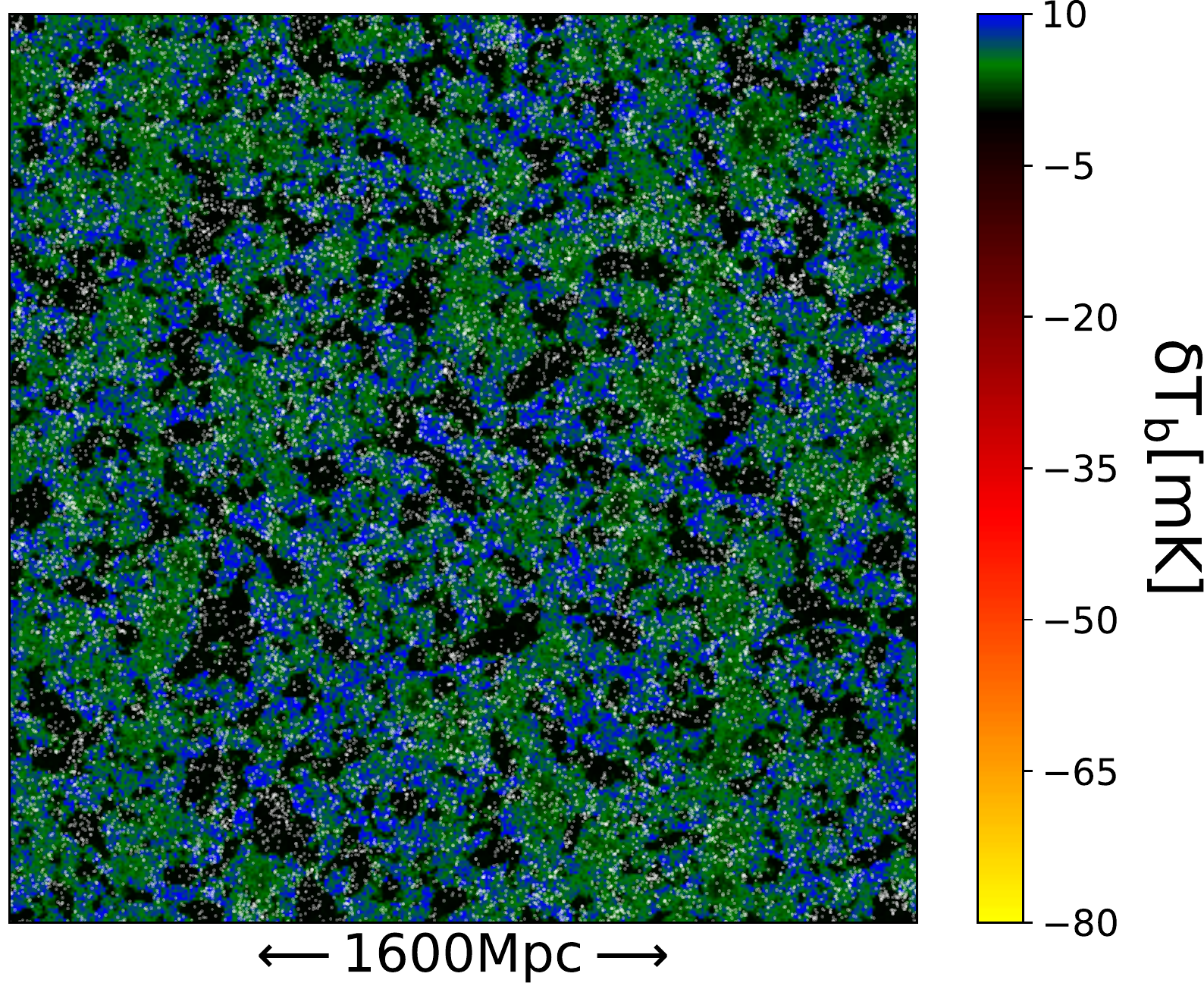}
\includegraphics[width=\columnwidth]{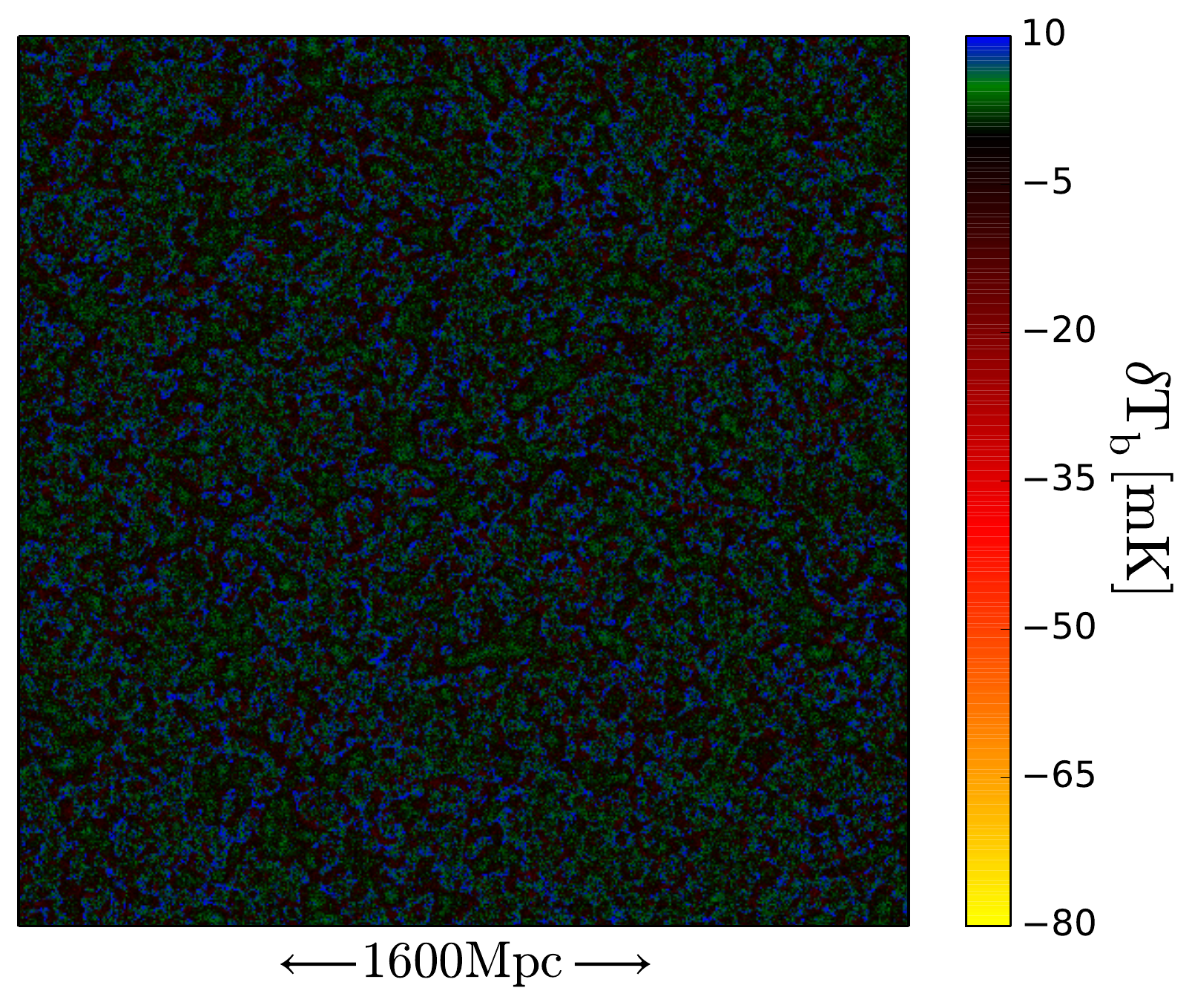}
\includegraphics[width=\columnwidth]{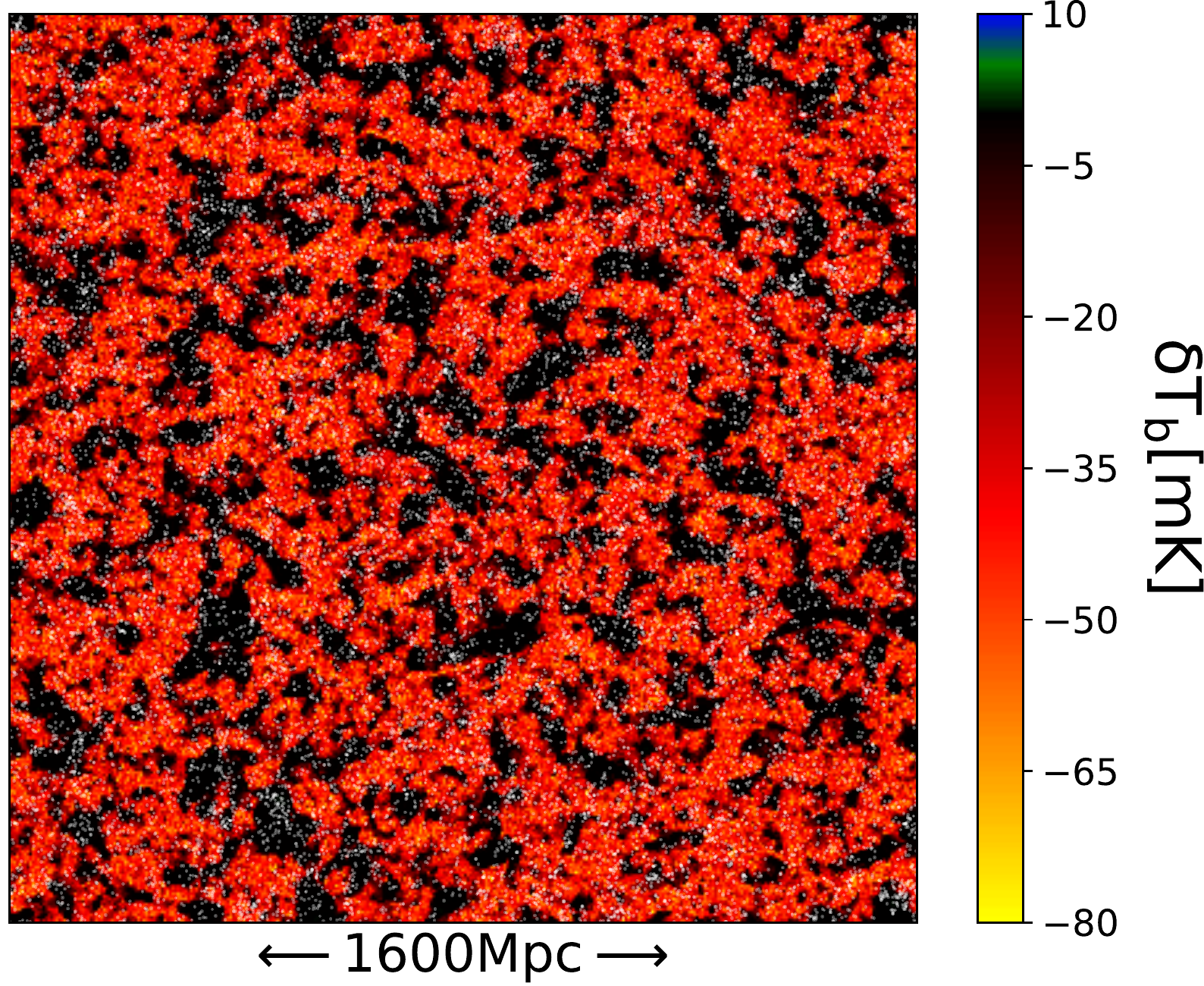}
\includegraphics[width=\columnwidth]{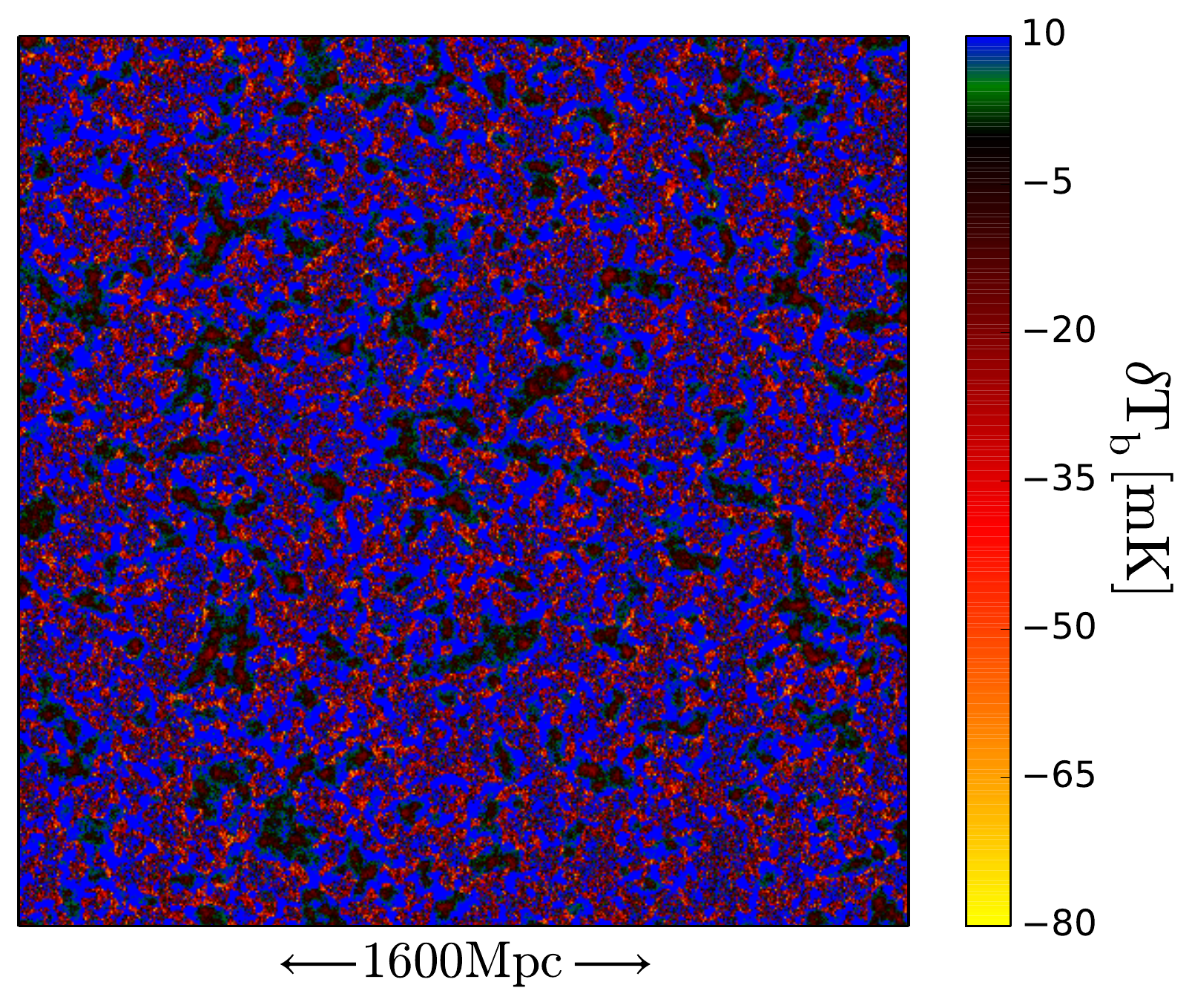}
\caption{{\it Left column}: 21-cm brightness temperature maps at $\bar{x}_{\rm HI}=0.5$ from the LargeHII EOS simulation, together with LAEs (white dots), lying in a $\Delta z=$0.1 slice (this corresponds to roughly 38$\,$Mpc at $z\sim 6.6$).  {\it Right column}: Corresponding 21-cm maps, with SKA1-low 1000h noise realizations.  The top, middle, bottom rows correspond to an average spin temperature of $\aveTs$ =  640, 40, 6 K and an average brightness temperature of $\bar{\delta T}_{\rm b}$ 9, 4, -26 mK, respectively. The middle row corresponds to the fiducial values in the EOS LargeHII simulation.
  }
\label{FIG:mocks21}
\end{figure*}

\subsubsection{Mock SKA1-Low observations}\label{sec:mock21}

Our fiducial 21-cm observations correspond to an SKA1-low tracked scan with 1000h on-sky integration.
The associated noise is calculated using the publicly available 21cmSense code\footnote{\url{https://github.com/jpober/21cmSense}} \citep{Pober13, Pober14}.
Specifically, we apply the moderate foreground option in 21cmSense, which assumes modes in the so-called foreground wedge are lost (i.e. have infinite noise). We assume 6 hours per night tracked scan and 167 days per year. Frequency-dependent scaling for sky temperature is assumed to follow $T_\mathrm{sky}=60\lambda^{2.55}$ with wavelength $\lambda$~\citep{TMS07}. The system temperature is therefore frequency-dependent, following the relation $T_\mathrm{rec} = 1.1 T_\mathrm{sky} + 40$mK as outlined in the SKA System Baseline Design.\footnote{\url{https://www.skatelescope.org/wp-content/uploads/2012/07/SKA-TEL-SKO-DD-001-1\_BaselineDesign1.pdf}} We use the configuration from the SKA1-low baseline design with a compact antennae core that has a maximal baseline of 1.7 km (longer and more sparsely sampled baselines are mainly for calibration purposes and add little sensitivity to the EoR signal).

We generate mock 21-cm observations by randomly sampling the above mentioned noise power in Fourier space, and adding it to the cosmological signal. For the cosmological signal we use the EOS simulations adjusted to $z=6.6$ as described in \S~\ref{sec:21cm}.

In Figure~\ref{FIG:mocks21} we show the simulation boxes alongside the generated mock signal for three different assumptions on the spin temperature at $x_\mathrm{HII}\sim 0.5$.  The panels correspond to the post-heating regime, the fiducial spin temperature field of the LargeHII EOS simulation, and pre-heating when the IGM was on average colder than the CMB, from top to bottom.  For samples in our ($\avenf$, $\aveTs$) parameter space, we compute 10 such noise realizations which are used to Monte Carlo sample the expected scatter in the cross-correlation (see \S~\ref{sec:rcorr}).  

\subsection{Cross-correlation statistics}
Throughout this paper we use as a statistic for the cross-signal between the 21cm signal and the LAE maps the real-space cross-correlation function $r_\mathrm{21,LAE} \left( r \right) \equiv \langle \delta_{21}\left(x\right) \delta_\mathrm{LAE}\left(x+r\right)\rangle_\mathrm{x}$.  Alternately, one could work with the Fourier equivalent, the cross power, but the two present similar trends (e.g. \citealt{Vrbanec16, SMG16}); the real-space cross-correlation function has the added benefit of having a more physical normalization and interpretation as an excess probability compared to random.

In the above expression, the LAE overdensity is:
\begin{equation}
\delta_\mathrm{LAE}({\bf x}) = \frac{N_\mathrm{LAE}({\bf x})}{\bar{N}_\mathrm{LAE}} - 1 ,
\end{equation}
with $N_\mathrm{LAE}({\bf x})$ denoting the number of  LAE in a voxel at position {\bf x}, and $\bar{N}_\mathrm{LAE}$ corresponding to the mean that is kept constant to match the observed Subaru HSC number density, as discussed previously.
The brightness temperature fluctuations are defined as
\begin{equation}
\delta_{21}({\bf x}) = \frac{T_{21} ({\bf x}) - \bar{T}_{21}}{T_0} ,
\end{equation}
where $T_0=23.5$ mK  
is the expected brightness temperature at $z=6.6$ for an entirely neutral universe and $\bar{T}_{21}$ is the actual mean brightness temperature. Furthermore, $P_\mathrm{21,LAE} \equiv k^3/(2\pi^2 V ) \mathfrak{R} \langle 
\delta_{21}\delta_\mathrm{LAE}\rangle_\mathrm{k}$ is the cross-power spectrum between the 21cm-signal and LAEs. 

In practice we calculate the cross-correlation function directly from our real-space\footnote{In principle we could also directly Fourier transform from the cross-power spectrum to the cross-correlation function; however, we find that the direct real-space calculation is more stable in the presence of 21-cm noise.}
21-cm boxes and mock LAE catalogues for the same 3.5 deg$^2$ field.  We
follow the metric from~\citet{Croft16}, summing over the visible, 2D projected galaxy--21cm pixel pairs that are separated by distance $r$, 
\begin{equation}
r_\mathrm{21,LAE} \left( r \right) = \frac{1}{N_\mathrm{LAE,s}N(r)} \sum_i^{N_\mathrm{LAE,s}} \sum_j^{N(r)} \delta_{21}\left(\textbf{r}_{\rm i}+\textbf{r}_{\rm j}\right)~,
\label{eq:r21LAE}
\end{equation}
where $\textbf{r}_{\rm i}$ is the position of the $i$-th LAE and $|\textbf{r}_{\rm j}|=r$; $N_{\rm gal}$ is the number of LAEs in the survey and $N\left(r\right)$ is the number of pixels at distance $r$ from the i-th LAE.
To quantify the uncertainty on the cross-correlation, we compute eq. (\ref{eq:r21LAE}) with 10 different Monte Carlo realizations of the SKA1-low noise for every 21-cm map used (c.f. \citealt{Vrbanec16}).

\section{Results}\label{sec:cross}

\subsection{21cm -- LAE cross-correlation: general trends}\label{sec:rcorr}

\begin{figure}
\centering 
\includegraphics[width=0.9\columnwidth]{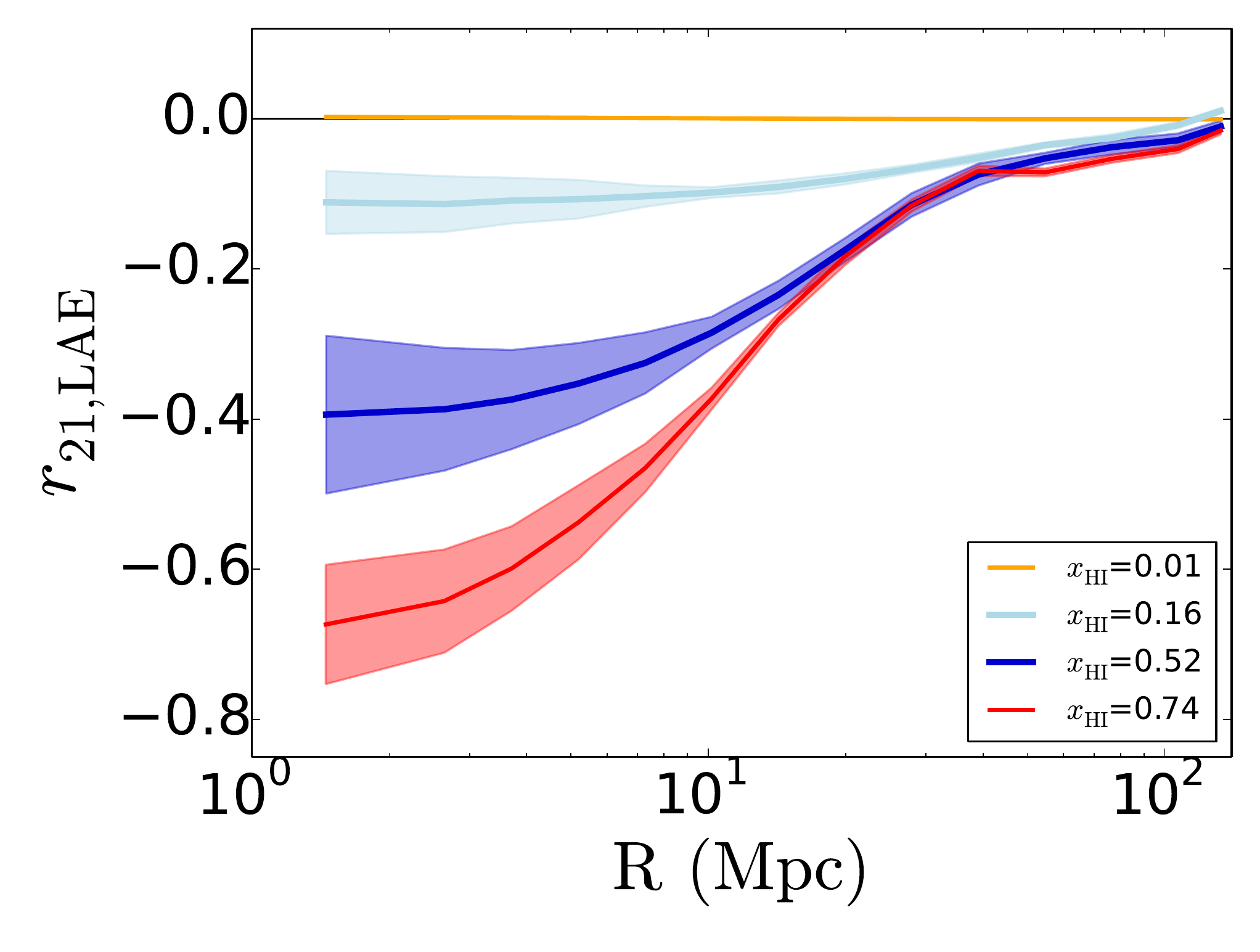}
\includegraphics[width=0.9\columnwidth]{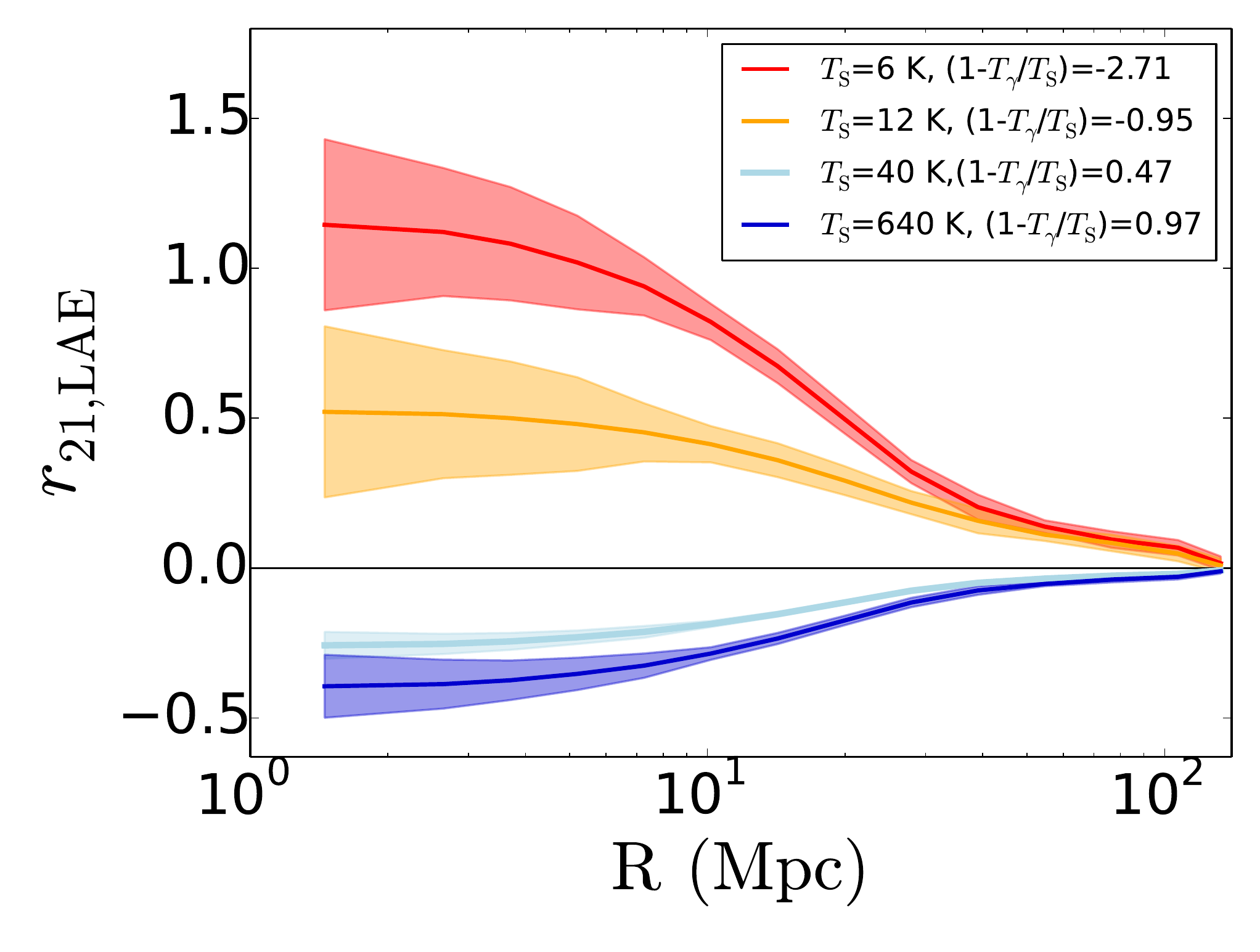}
\caption{21cm -- LAE correlation functions, assuming the LargeHII reionization topology from EOS. Shaded regions indicate 2$\sigma$ scatter computed from 10 mock SKA1-low + Subaru HSC observations as described in the text. In the top panel, we vary the neutral fraction, assuming high $T_\mathrm{S}$ with $\left(1-T_{\gamma}/T_\mathrm{S} \right)>0.9$ close to the usual saturated spin temperature limit.  In the bottom panel, we fix $\avenf = 0.5$, and instead vary the spin temperature.  The cross-correlation is always less negative compared to the saturated limit.
}
\label{FIG:rcorr}
\end{figure}

We begin by illustrating in Figure~\ref{FIG:rcorr}  some general trends of the 21cm -- LAE cross-correlation, as we vary ($\avenf$, $\aveTs$) at $z=6.6$ for the LargeHII reionization morphology.  In the top panel, we show how the cross-correlation varies as a function of the neutral fraction, taking $\left(1- T_{\rm CMB}\ T_{\rm S}\right) > 0.9$ to approximate the saturated spin temperature limit $\aveTs \gg T_{\rm CMB}$. This is the assumption made by previous studies (e.g. \citealt{Vrbanec16, SMG16,  HTD18, Kubota19}), and we recover their result that the cross-correlation is in general negative.  This is because cosmic reionization (pre-overlap) is inside-out: large-scale overdensities which contain more galaxies ionize before those having fewer galaxies (e.g. \citealt{TG11}).  Thus inside HII regions we have $\delta_{21} < 0$ (zero brightness temperature, which is less than the mean) and $\delta_{\rm LAE} > 0$ (overdensity of galaxies), while in the neutral regions we have $\delta_{21} > 0$ (positive brightness temperature) and $\delta_{\rm LAE} < 0$ (underdensity of galaxies).  This cross-correlation becomes more negative with increasing $\avenf$, since the bias of the cosmic HII regions correspondingly increases (the rarest, most biased galaxies were the first to ionize their surroundings).  Moreover, the characteristic scale (e.g. when the correlation function is half of the maximum amplitude), also decreases with increasing $\avenf$ as the cosmic HII regions become smaller.

However, this is no longer true if the neutral IGM is colder than the CMB, and is thus seen in absorption.  In this case, the neutral IGM has a {\it negative} brightness temperature while the ionized IGM still has a zero brighness temperature.
Thus when the 21cm is seen in absorption, the cosmic HII regions which have an overabundance of galaxies become relative hot spots in 21cm, and so the 21cm --LAE cross-correlation becomes {\it positive}.\footnote{Here we present forecasts for cross-correlation with LAEs, but this trend holds for cross-correlating 21cm with any tracer of the matter field (e.g. \citealt{Moriwaki19}).}

We see this very trend in the bottom panel of Figure~\ref{FIG:rcorr}, where we fix $\avenf = 0.5$ and instead vary $\aveTs$.  The cross-correlation switches to being positive roughly when $(1- T_{\rm CMB}/\aveTs)$ becomes negative.  Even when the spin temperature is higher than the CMB and 21cm is seen in emission, the strength of the negative cross-correlation is generally less than the saturated limit implies.  This is because assuming a saturated spin temperature gives the {\it maximum} achievable contrast between the ionized and neutral IGM.
For example, if $\aveTs = 2 T_{\rm CMB}$, the neutral regions would have a brightness temperature $(1- T_{\rm CMB}/\aveTs) = 0.5$ times smaller than in the case $\aveTs \gg T_{\rm CMB}$.  Hence the contrast between the ionized and neutral regions in 21cm is decreased.

Regarding the detectability of this cross-signal, an instrument like SKA1-low cross-correlated with a LAE survey similar to Subaru/HSC should be able to detect the 21cm -- LAE cross-signal for all of the models shown. We caution however than although the different models are distinguishable within the errors, here we only show results for a single EoR morphology: LargeHII from EOS.  The actual EoR morphology is unknown, as it depends which sources (i.e. their bias) are the dominant emitters of ionizing photons (e.g. \citealt{McQuinn07}).  As shown in \citet{SMG16}, the EoR morphology is the largest uncertainty in the 21cm -- LAE cross-correlation forecasts.  Therefore although a high S/N detection is very feasible, actually inferring IGM properties through the cross-signal will involve assumptions on the EoR topology/galaxy modeling.  Below we illustrate this further, by showing analogous measurements for both the SmallHII and LargeHII morphologies.

\subsection{21cm -- LAE cross-correlation: exploration of ($\bar{x}_\mathrm{HI}$, $\aveTs$) parameter space}\label{sec:spin}

We now explore the full parameter space dependence of the 21cm -- LAE cross-signal, co-varying both the spin temperature $\bar{T}_\mathrm{S}$ and the neutral fraction $\bar{x}_\mathrm{HI}$.  We  also show results for different EoR and epoch of X-ray heating (EoH) morphologies.

\begin{figure*}
\centering 
\begin{minipage}{.45\textwidth}
    \centering
   \includegraphics[width=\columnwidth]{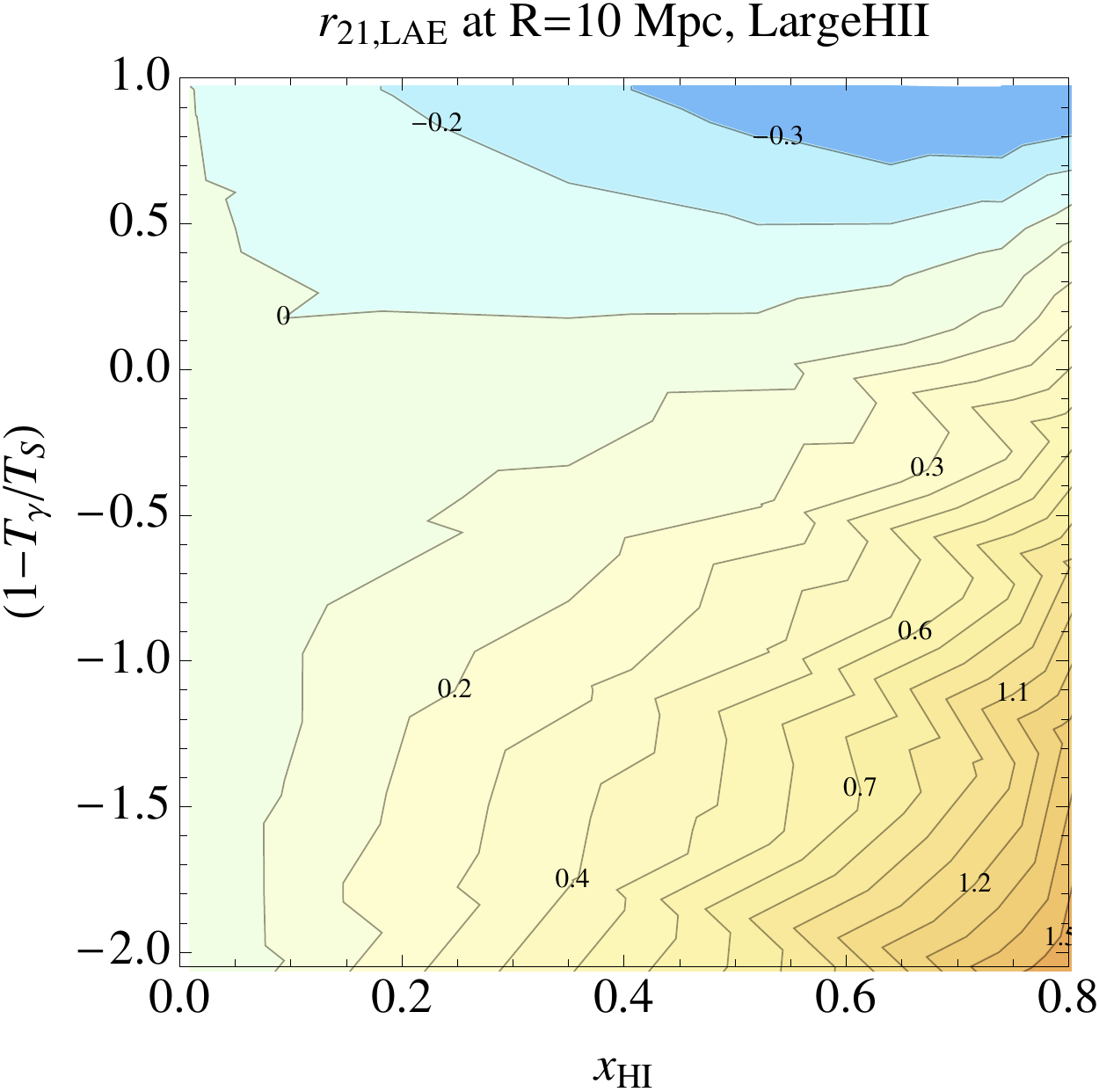}
    \centering
   \includegraphics[width=\columnwidth]{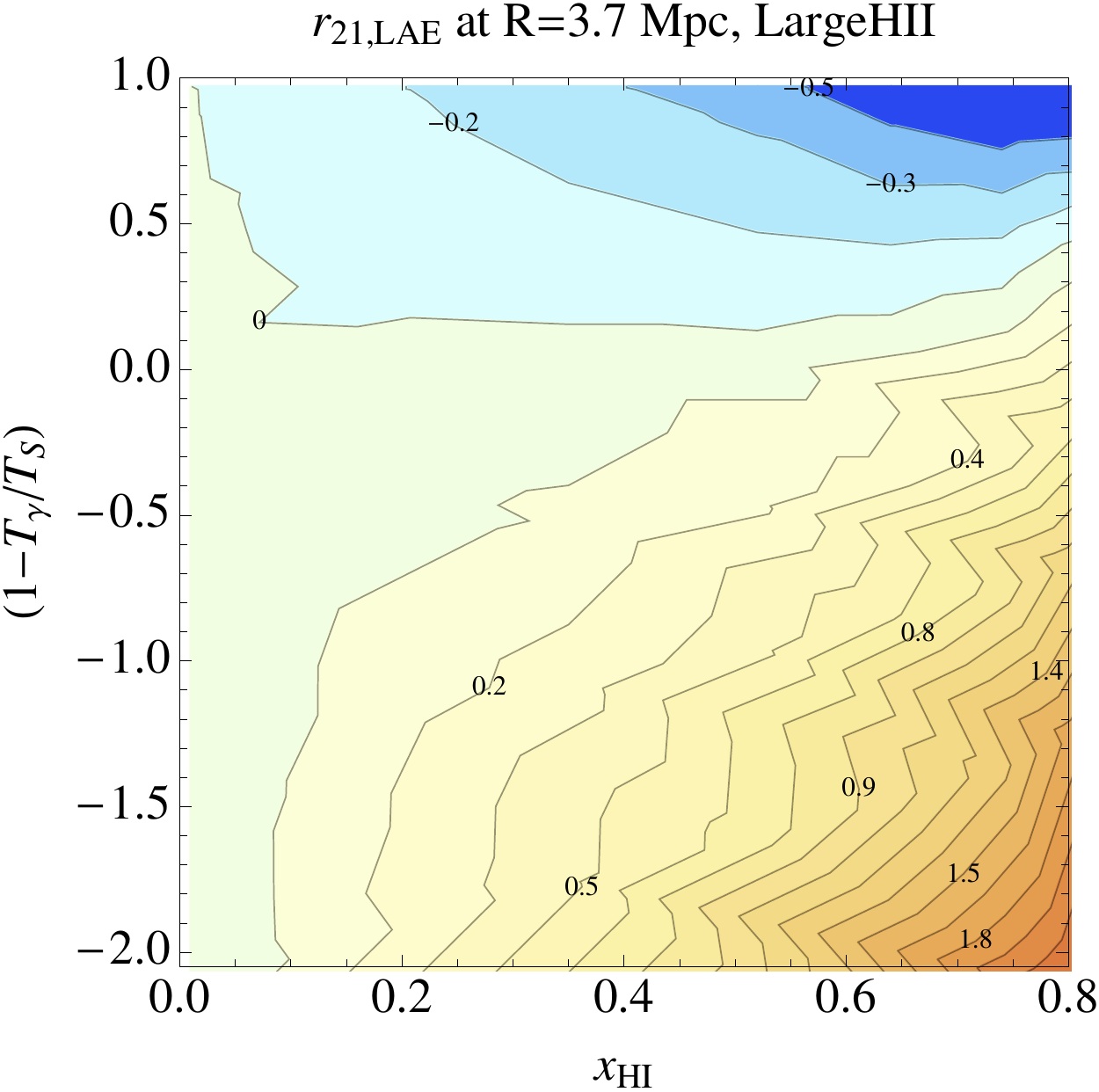}
\end{minipage}%
\begin{minipage}{.45\textwidth}
    \centering
   \includegraphics[width=\columnwidth]{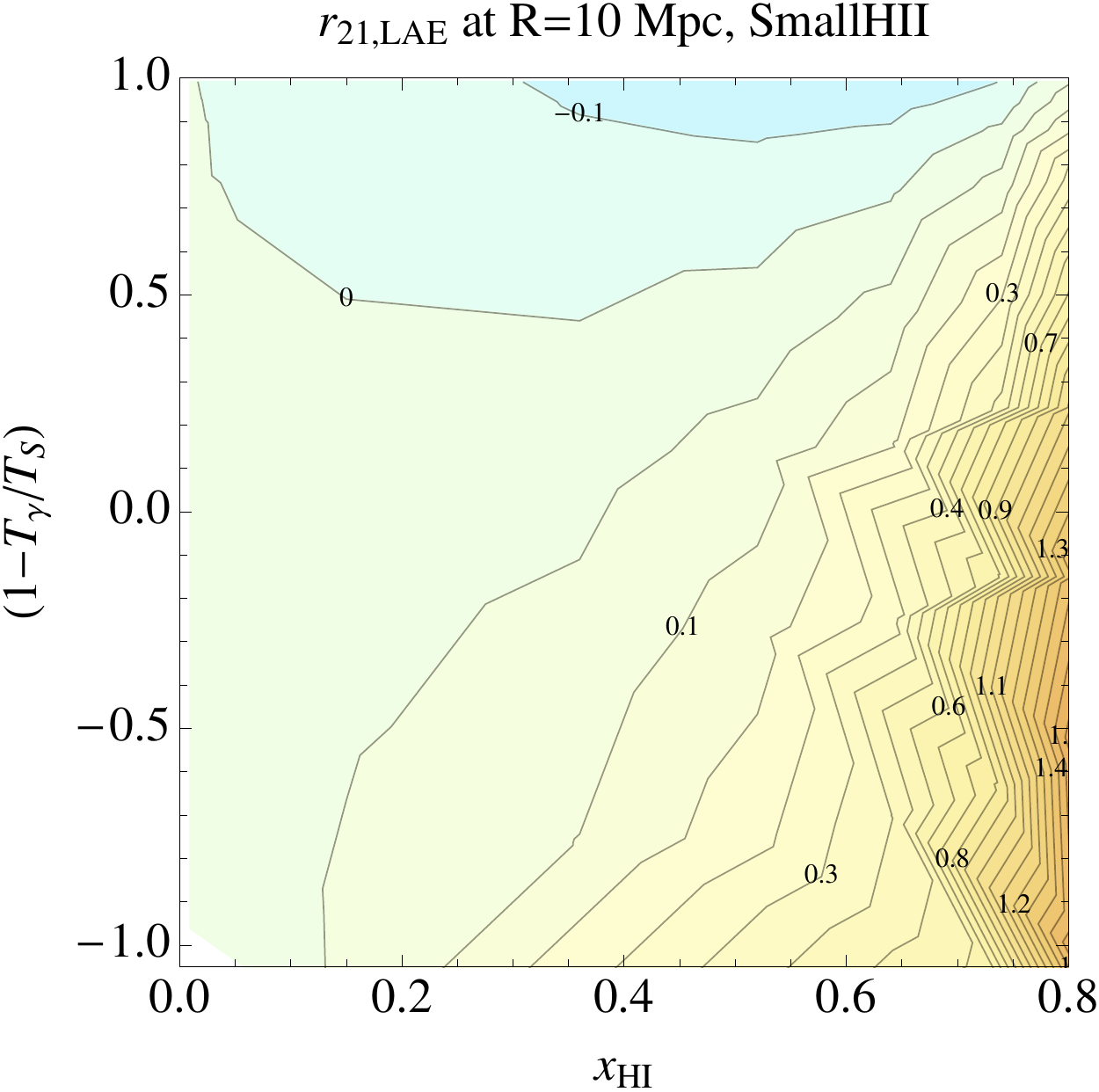}
    \centering
   \includegraphics[width=\columnwidth]{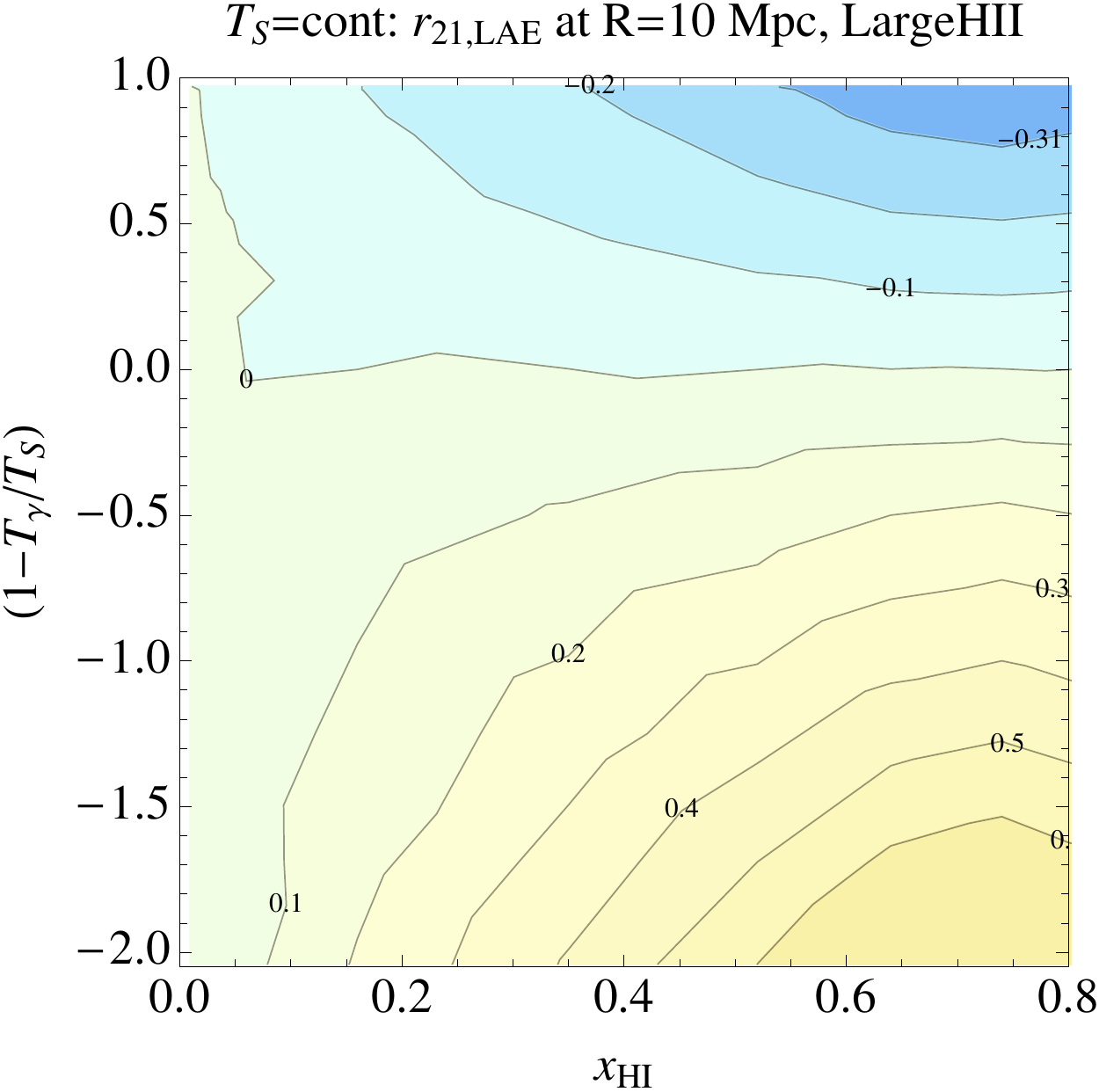}
\end{minipage}%
\begin{minipage}{.1\textwidth}
    \centering
    \includegraphics[width=0.9\columnwidth]{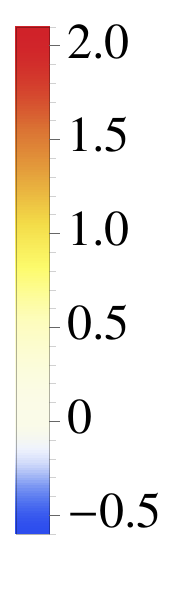}
\end{minipage}
\caption{Contours of the 21cm -- LAE cross-correlation, as a function of $\avenf$ and $\aveTs$ at $z=6.6$ for our mock surveys.  {\it Top left:} The cross-correlation function evaluated at R=10$\,$Mpc, for the LargeHII model. {\it Top right:} the same as top left, but for the SmallHII model. {\it Bottom left:} the same as the top left, but evaluated at $R$=3.7$\,$Mpc. {\it Bottom right:} the same as top left, but assuming a homogeneous spin temperature in the neutral IGM, instead of the inhomogeneous X-ray heating tracked by the simulation. 
}
\label{FIG:rTs}
\end{figure*}

In the top left panel of Figure~\ref{FIG:rTs}, we plot isocontours of the 21cm -- LAE cross-correlation function for the LargeHII model, evaluated at $R=10$ Mpc.  The choice of scale ensures high S/N measurements, and is a factor of several larger than our resolution of 1.6 Mpc.  As the cross-correlation functions are smooth and featureless, the general trends inferred at $R=10$ Mpc should hold for other scales, as we explicitly verify below.

Post-reionization ($\avenf \rightarrow 0$), we note a very small positive cross-correlation, since the 21-cm signal in the post-overlap regime is sourced by the residual HI inside LLSs and DLAs.  These relatively dense structures preferentially reside in the same large-scale matter overdensities as do galaxies (e.g. \citealt{MHR00, FO05, POW10, Munoz16}).

At a fixed neutral fraction, the amplitude of the cross-correlation tends to increase with the absolute value of $\left| 1 - T_{\rm CMB}/T_{\rm S} \right|$. The positive cross-correlation is able to reach much larger amplitudes than the negative one. This is because there is a larger dynamic range available for the brightness temperature, through the $(1 - T_{\rm CMB}/T_{\rm S})$ term, when the signal is in absorption, $T_{\rm S} < T_{\rm CMB}$. Indeed this is also the reason why the highest peak in the 21-cm auto power spectrum is expected to be during the EoH, when the signal is seen in absorption against the CMB (e.g. \citealt{PF07, MFS13}).
At a fixed spin temperature, the amplitude of this positive cross-correlation increases with increasing neutral fraction.  As mentioned previously, this is again due to the increasing bias of the cosmic HII regions.  Outside of the local cosmic HII regions, the IGM is preferentially hotter, heated by the X-rays from the first galaxies.  Thus the environments of these galaxies always correspond to the high value tail of the brightness temperature distribution, when the bulk of the IGM is cold.

In the top right panel we show the analogous plot, but instead computed with the SmallHII EOS model.  We note the same qualitative trends as seen for the LargeHII model for the EoR/EoH morphology.  Comparing the two morphologies at the same points in ($\avenf$, $\aveTs$) parameter space, we note that the amplitude of the cross-correlation is somewhat smaller in the SmallHII model.  Having smaller, more evenly distributed 21-cm structures results in a smaller 21-cm auto power spectrum (e.g. \citealt{McQuinn07}), and a correspondingly smaller 21cm -- LAE cross-correlation \citep{SMG16}. Relatedly, the cross-correlation remains positive at high values of $\avenf$, even when the bulk of the IGM is in emission.  Here, the large-scale positive correlation of both LAE and 21cm with the underlying matter field comes through, as the small, disjoint HII regions do not mask out the peaks of 21cm emission.

In the bottom left panel we show the analogous plot as in the top left, however evaluated instead at $R=3.7$ Mpc instead of $R=10$ Mpc.  Qualitatively, there is very little difference between the two panels, highlighting that the cross-correlation is fairly featureless over this range of scales.  Only the amplitudes are somewhat larger at these smaller scales.

Finally, in the bottom right panel, we show the analogous plot as in the top left;  however, instead of taking the inhomogeneous spin temperature maps computed by the EOS simulation, here we set the neutral IGM to have a {\it uniform} value for the spin temperature. 
Self-consistently computing X-ray heating and Lyman alpha coupling is computationally challenging (e.g. \citealt{Baek10, Santos11, MFC11, Eide18, Ross18}), and requires very large box sizes ($\gtrsim$ 250 Mpc; see \citealt{2020arXiv200406709D}).  Being able to ignore temperature fluctuations would simplify the cross-correlation calculation dramatically (e.g. \citealt{Pober15}).

Comparing the top left and bottom right panels, we see that the spin temperature inhomogeneity is not as important in the late stages of reionization, i.e. the contours are similar at $\avenf \lesssim 0.5$.  However, in the early stages of the EoR and EoH, temperature fluctuations have a substantial impact.  Specifically, cross-correlation in the uniform temperature model is fairly independent of the neutral fraction, missing the aforementioned increase in amplitude due to the increasing bias of the HII regions hosting the LAEs.  Thus the value of the cross-correlation in the early stages of the EoR can be used to probe the morphology of both the EoR and EoH.

\section{Summary and Conclusions}\label{sec:end}

Cross-correlation of 21cm and LAE can serve as an important proof of the cosmic origin of future 21cm detections.  Previous forecasts for this cross-correlation assumed that the spin temperature in the neutral IGM was much higher than the CMB temperature.  More recent calculations, using star formation rate densities infered from galaxy LFs, suggest that this assumption is unlikely to be true for a large part of reionization.

Here we revisit the 21cm - LAE cross-correlation, relaxing the assumption of a pre-heated IGM before reionization.  We make mock forecasts over a range of mean IGM neutral fractions and the mean spin temperature of the neutral IGM, using SKA1-low and Subaru HSC specifications.

We show that the real-space cross-correlation function is strongly dependent on both $\avenf$ and $\aveTs$.  If the IGM is seen in emission against the CMB, the cross-correlation is generally negative, since the ionized regions hosting galaxies ($\Delta T \sim 0$ mK) correspond to relative cold spots in the 21-cm brightness temperature field.   If the IGM is seen in absorption against the CMB, the cross-correlation is generally positive.  In this case, the ionized regions hosting galaxies correspond to relative hot spots in the brightness temperature field.

We also vary the topology of the EoR as well as the EoH.  We show that the cross-correlation during the second half of reionization is fairly insensitive to the EoR and EoH morphologies, when compared at a given ($\avenf$, $\aveTs$).  However during the early stages of reionization, the topology does impact both the amplitude and the sign of the cross-correlation.  Thus the 21cm -- LAE can tell us about the typical galaxy populations whose UV and X-ray radiation drives the signal.

\section*{Acknowledgements}
We are extremely grateful to M. Ouchi for providing us with HSC survey data, and to Bradley Greig for his help with 21cmSense and its setup for SKA.
This project receives funding from the European Research
Council (ERC) under the European Union's Horizon 2020 research and innovation programme (grant agreement No 638809 -- AIDA).  The ERC is not responsible for the views presented here.




\bibliographystyle{mnras}
\bibliography{ms} 




\bsp	
\label{lastpage}
\end{document}